\documentstyle[amssymb,preprint,aps,epsfig]{revtex}

\begin{document}
\draft
\title{Permutation zones and the fermion sign problem}
\author{Dean Lee}
\address{Dept. of Physics, North Carolina State University, Raleigh, NC 27695}
\maketitle

\begin{abstract}
We present a new approach to the problem of alternating signs for fermionic
many body Monte Carlo simulations. \ We demonstrate that the exchange of
identical fermions is typically short-ranged even when the underlying
physics is dominated by long distance correlations. \ We show that the
exchange process has a maximum characteristic range of $\sqrt{2(1-f)\beta h}$
lattice sites$,$ where $\beta $ is the inverse temperature, $h$ is the
hopping parameter, and $f$ is the filling fraction. \ We introduce the
notion of permutation zones, special regions of the lattice where identical
fermions may interchange and outside of which they may not. $\ $Using
successively larger permutation zones, one can extrapolate to obtain
thermodynamic observables in regimes where direct simulation is impossible.
\end{abstract}


\section{Introduction}

One of the most important challenges for quantum field theory and many body
simulations is the study of light fermion dynamics. \ Today most fermion
algorithms used in lattice field theory are based on pseudofermion methods,
which calculate the contribution of fermions indirectly by means of an
effective non-local bosonic action. The most popular approach, Hybrid Monte
Carlo, deals directly with non-local actions \cite{hmc}. \ Molecular
dynamics is used to propose new configurations and a Metropolis criterion is
used to accept or reject updates. \ Although the advantages of pseudofermion
methods are substantial, there are still important motivations for doing
simulations with explicit fermions. \ The primary concern is the question of
what fermions actually do in lattice simulations. \ In lattice gauge theory,
for example, one might be interested in seeing the effects of confinement on
quark-antiquark pair separation or the relation between chiral symmetry
breaking and light quark mobility. \ These questions are most easily
addressed in a local field framework which keeps the fermionic degrees of
freedom explicit.

In order to do get anywhere with explicit fermion simulations one must of
course address the fermion sign problem. In this paper we introduce a new
approach to the sign problem which has applications to quantum simulations
at finite temperature. \ Unlike the fixed-node approach \cite
{anderson,reynolds,moskowitz}, our method makes no assumption about the
nodal structure of the eigenfunctions. \ It also is not a resummation
technique, the underlying principle powering the meron-cluster algorithm 
\cite{chandrasekharan} and diagonalization/Monte Carlo methods \cite
{lee1,lee2}. \ The approach we introduce here is based on the observation
that in most finite temperature simulations fermion permutations are short
ranged. \ This holds true even for systems with massless modes and long
distance correlations, as we demonstrate with two examples. \ In this paper
we focus on the application of the zone method to simulations with explicit
fermions. \ These methods have recently been applied to study chiral
symmetry breaking in massless quantum electrodynamics in $2+1$ dimensions 
\cite{maris}. \ The extension to pseudofermion methods and, in particular,
applications to Euclidean lattice gauge theory will be discussed in a future
publication.

\section{Worldlines}

We begin with a brief review of the worldline formalism \cite{hirsch}. \ We
introduce the basic ideas in one spatial dimension before moving on to
higher dimensions. \ Let us consider a system with one species of fermion on
a periodic chain with $L$ sites, where $L$ is even. \ Aside from an additive
constant, the general Hamiltonian can be written as 
\begin{equation}
H=-h\sum_{i}\left[ a_{i+1}^{\dagger }a_{i}+a_{i}^{\dagger }a_{i+1}\right]
+\sum_{i}c_{i}a_{i}^{\dagger }a_{i}.
\end{equation}
Following \cite{hirsch} we break the Hamiltonian into two parts, $H_{\text{e}%
}$ and $H_{\text{o}}$, 
\begin{equation}
H_{\text{e/o}}=\sum_{i\text{ even/odd}}\left[ 
\begin{array}{c}
-ha_{i+1}^{\dagger }a_{i}-ha_{i}^{\dagger }a_{i+1} \\ 
+%
{\textstyle{c_{i} \over 2}}%
a_{i}^{\dagger }a_{i}+%
{\textstyle{c_{i+1} \over 2}}%
a_{i+1}^{\dagger }a_{i+1}
\end{array}
\right] .
\end{equation}
We note that $H=H_{\text{e}}+H_{\text{o}}$.

We are interested in calculating thermal averages, 
\begin{equation}
\left\langle O\right\rangle =\frac{Tr\left[ O(a^{\dagger },a)\exp (-\beta H)%
\right] }{Tr\left[ \exp (-\beta H)\right] },
\end{equation}
where $\beta =(k_{B}T)^{-1}$. \ For large $N$, we can write 
\begin{equation}
\exp (-\beta H)=\left[ \exp \left[ -%
{\textstyle{\beta  \over N}}%
(H_{\text{e}}+H_{\text{o}})\right] \right] ^{N}\approx (S_{\text{o}}S_{\text{%
e}})^{N},  \label{exp}
\end{equation}
where 
\begin{equation}
S_{\text{e/o}}=\exp (-%
{\textstyle{\beta  \over N}}%
H_{\text{e/o}}).
\end{equation}
Inserting a complete set of states at each step, we can write $Tr\left[ \exp
(-\beta H)\right] $ as 
\begin{equation}
\sum_{z_{0},...,z_{2N-1}}\left\langle z_{0}\right| S_{\text{o}}\left|
z_{2N-1}\right\rangle ...\left\langle z_{1}\right| S_{\text{e}}\left|
z_{0}\right\rangle .  \label{b}
\end{equation}

In Fig. 1 a typical set of states $\left| z_{0}\right\rangle ,...,\left|
z_{2N-1}\right\rangle $ are shown which contribute to the sum in (\ref{b}).\
\ We call such a contribution a worldline configuration. \ The shaded
plaquettes represent locations where $S_{\text{e}}$ or $S_{\text{o}}$ acts
on the corresponding local fermionic state. \ The classical trajectory of
each\ of the fermions can be traced from Euclidean time $t=0$ to time $%
t=\beta $. \ In the case when two identical fermions enter the same shaded
plaquette, we adopt the convention that the worldlines run parallel and do
not cross. \ With this convention the fermion sign associated with Fermi
statistics is easy to compute. \ The worldlines from $t=0$ to $t=\beta $
define a permutation of identical fermions. \ Even permutations carry a
fermion sign of $+1$ while odd permutations carry sign $-1$. \ The
generalization to higher dimensions is straightforward. \ In two dimensions,
for example, $\exp (-\beta H)$ takes the form 
\begin{equation}
\left. 
\begin{array}{c}
\left[ \exp \left[ -%
{\textstyle{\beta  \over N}}%
(H_{\text{e}}^{x}+H_{\text{o}}^{x}+H_{\text{e}}^{y}+H_{\text{o}}^{y})\right] 
\right] ^{N} \\ 
\qquad \qquad \qquad \approx (S_{\text{o}}^{y}S_{\text{e}}^{y}S_{\text{o}%
}^{x}S_{\text{e}}^{x})^{N}.
\end{array}
\right.  \label{2d}
\end{equation}

The sum over all worldline configurations can be calculated with the help of
the loop algorithm \cite{evertz}. \ At each occupied/unoccupied site, we
place an upward/downward pointing arrow as shown in Fig. 2. \ Due to fermion
number conservation, the number of arrows pointing into a plaquette equals
the number of arrows pointing out of the plaquette. \ New Monte Carlo
updates of the worldlines are produced by flipping the arrows which form
closed loops. \ 

\section{Wandering length}

In one spatial dimension, the Pauli exclusion principle inhibits fermion
permutations except in cases where the fermions wrap around the lattice
boundary. \ For the remainder of our discussion, therefore, we consider
systems with two or more dimensions. \ The first question we address is how
far fermion worldlines can wander from start time $t=0$ to end time $t=\beta 
$. \ We can put an upper bound on this wandering distance by considering the
special case with no on-site potential and only nearest neighbor hopping.

Let us consider motion in the $x$-direction. \ For each factor of $S_{\text{o%
}}^{x}S_{\text{e}}^{x}$ in (\ref{2d}) a given fermion may remain at the same 
$x$ value, move one lattice space to the left, or move one lattice space to
the right. \ If $h$ is the hopping parameter, then for large $N$ the
relative weights for these possibilities are approximately $1$ for remaining
at the same $x$ value, $\beta hN^{-1}$ for one move to the left, and $\beta
hN^{-1}$ for one move to the right. \ In (\ref{2d}) we see that there are $N$
factors of $S_{\text{o}}^{x}S_{\text{e}}^{x}$. \ Therefore for a typical
worldline configuration at low filling fraction, $f,$ we expect $\sim\beta h$
hops to the left and $\sim\beta h$ hops to the right. \ For non-negligible $%
f $ some of the hops are forbidden by the exclusion principle. \ Assuming
random filling we expect $\sim(1-f)\beta h$ hops to the left and $%
\sim(1-f)\beta h$ hops to the right.

The net displacement is equivalent to a random walk with $2(1-f)\beta h$
steps. \ The expected wandering length, $\Delta ,$ is therefore given by 
\begin{equation}
\Delta =\sqrt{2(1-f)\beta h}.  \label{wandering}
\end{equation}
This result is somewhat surprising in that for typical simulation parameters
(i.e., $\beta $ not too large), we find $\Delta $ is no larger than a few
lattice units. \ There is no contradiction between the existence of long
distance correlations and the constraint of short distance wandering
lengths. \ Long range signals are propagated by the net effect of many short
range displacements. \ A simple analogy can be made with electrical
conduction in a wire or sound propagation in a gas, which results from many
short range displacements of individual electrons or gas molecules.

In cases with on-site potentials, fermion hopping is dampened by differences
in potential energy. \ Hence the estimate ($\ref{wandering}$) serves as an
upper bound for the general case. \ We have checked the upper bound
numerically using simulation data generated by several different lattice
Hamiltonians with and without on-site potentials.

\section{Permutation Zone Method}

Let $W$ be the logarithm of the partition function, 
\begin{equation}
W=\log \left\{ Tr\left[ \exp (-\beta H)\right] \right\} .
\end{equation}
Let us partition the spatial lattice, $\Gamma $, into zones $%
Z_{1},Z_{2},...,Z_{k}$ such that the spatial dimensions of each zone are
much greater than $\Delta $. \ For notational convenience we define $%
Z_{0}=\emptyset $. \ For any $R\subset \Gamma $, let $W_{R}$ be the
logarithm of a restricted partition function that includes only worldline
configurations where any worldline starting outside of $R$ at $t=0$ returns
to the same point at $t=\beta $. \ In other words there are no permutations
for worldlines starting outside of $R$. We note that $W_{\Gamma }=W,$ and $%
W_{\emptyset }$ is the logarithm of the restricted partition function with
no worldline permutations at all. \ Since the zones are much larger than the
length scale $\Delta $, the worldline permutations in one zone has little or
no effect on the worldline permutations in another zones. \ Therefore 
\begin{equation}
W_{Z_{0}\cup ...\cup Z_{j}}-W_{Z_{0}\cup ...\cup Z_{j-1}}\approx
W_{Z_{j}}-W_{\emptyset }.
\end{equation}
Using a telescoping series, we obtain 
\begin{eqnarray}
W_{\Gamma }-W_{\emptyset } &=&\sum_{j=1,...,k}(W_{Z_{0}\cup ...\cup
Z_{j}}-W_{Z_{0}\cup ...\cup Z_{j-1}}) \\
&\approx &\sum_{j=1,...,k}(W_{Z_{j}}-W_{\emptyset }).  \nonumber
\end{eqnarray}
For translationally invariant systems tiled with congruent zones we find 
\begin{equation}
W=W_{\Gamma }\approx W_{\emptyset }+%
{\textstyle{\left| \Gamma \right|  \over \left| Z_{1}\right| }}%
(W_{Z_{1}}-W_{\emptyset }),  \label{tile}
\end{equation}
where $\left| \Gamma \right| /\left| Z_{1}\right| =k$, the number of zones.
\ For general zone shapes one can imagine partitioning the zones themselves
into smaller congruent tiles. \ Therefore the result (\ref{tile}) should
hold for large arbitrarily shaped zones. \ For this case we take $\left|
\Gamma \right| $ to be the number of nearest neighbor bonds in the entire
lattice and $\left| Z_{1}\right| $ to be the number of nearest neighbor
bonds in the zone. \ We will refer to $\left| Z_{1}\right| $ as the zone
size of $Z_{1}$. \ This is just one choice for zone extrapolation. \ A more
precise and complicated scheme could be devised which takes into account the
circumscribed volume, number of included lattice points, and other geometric
quantities.

\section{Free fermions}

As an example of the zone method, we compute the average energy $%
\left\langle E\right\rangle h^{-1}$ for a free fermion Hamiltonian with only
hopping interactions on an $8\times 8$ lattice. \ We consider values $\beta
h=1.0$, $1.5$, and $2.0$. \ The corresponding values for $\Delta $ are $1.0$%
, $1.2$, and $1.4$ respectively. \ The Monte Carlo updates are performed
using a single loop flip version of the loop algorithm \cite{evertz}.

In Fig. 3 we show data for rectangular zones with side dimensions $0\times 0$%
, $1\times 1$, $2\times 1$, $2\times 2$, $3\times 2$, $3\times 3$, ..., $%
6\times 6$. \ We also show a least-squares fit (not including the smallest
zones $0\times 0$ and $1\times 1$) assuming linear dependence on zone size
as predicted in (\ref{tile}). \ We find agreement at the $1\%$ level or
better when compared with the exact answers shown on the far right, which
were computed using momentum-space decomposition.

While the physics of the free hopping Hamiltonian is trivial, the
computational problems are in fact maximally difficult. \ The severity of
the sign problem can be measured in terms of the average sign, $<$Sign$>$,
for contributions to the partition function. \ For $\beta h=1.0$, $<$Sign$%
>\sim 0.005;$ for $\beta h=1.5$, $<$Sign$>$ $\sim 10^{-6};$ and for $\beta
h=2.0$, 
\mbox{$<$}%
Sign%
\mbox{$>$}%
$\sim 10^{-9}$. \ Direct calculation using position-space Monte Carlo is
impossible by several orders of magnitude for $\beta h\geq 1.5$.

\section{Fermionic 2D Ising Model}

While the free fermion example shows the linear dependence on zone size, we
now study an example which better demonstrates the utility of the zone
method. \ We will consider the Hamiltonian,

\begin{eqnarray}
H &=&J\sum_{i,j}\left[ s_{i,j}s_{i+1,j}+s_{i,j}s_{i,j+1}\right] 
\label{model} \\
&&-h\sum_{i,j}\left[ a_{i+1,j}^{\dagger }a_{i,j}+a_{i,j}^{\dagger
}a_{i+1,j}+a_{i,j+1}^{\dagger }a_{i,j}+a_{i,j}^{\dagger }a_{i,j+1}\right] , 
\nonumber
\end{eqnarray}
where 
\begin{equation}
s_{i,j}=2a_{i,j}^{\dagger }a_{i,j}-1=2n_{i,j}-1.
\end{equation}
We will refer to this model as the fermionic 2D Ising model. \ When $h=0$,
our model reduces to the antiferromagnetic 2D Ising model for $J>0$ and the
regular 2D Ising model for $J<0$, with $2n_{i,j}-1$ playing the role of
Ising spin. \ We also note that the model for $h>0$ is equivalent to the $h<0
$ model. \ On an even lattice, we can reverse the sign of $h$ by multiplying 
$-1$ to all creation and annihilation operators on sites where $i+j$ is odd.
\ Unlike other quantum generalizations such as the transverse field Ising
model (see for example \cite{sachdev}), we have reinterpreted the Ising spin
as an occupation number and introduced nearest neighbor hopping of
Fermi-Dirac particles.

To our knowledge the fermionic Ising model has not previously been discussed
in the literature. \ Therefore let us briefly discuss our interest in the
model. \ The model is motivated by our studies of time-dependent background
field fluctuations in Hamiltonian lattice gauge theories. \ In lattice gauge
theory we encounter time-dependent Hamiltonians which in the Kogut-Susskind
staggered formalism \cite{susskind} have the form 
\begin{equation}
H(t)=\sum_{i,j}\left[ H_{1,i,j}(t)+H_{2,i,j}(t)\right] +...,
\end{equation}
where 
\begin{eqnarray}
H_{1,i,j}(t) &=&\sum_{m,n}\left[ -(h_{1,i,j}^{m,n}(t))^{\ast
}a_{i+1,j}^{n\dagger }a_{i,j}^{m}-h_{1,i,j}^{m,n}(t)a_{i,j}^{m\dagger
}a_{i+1,j}^{n}\right] , \\
H_{2,i,j}(t) &=&\sum_{m,n}\left[ -(h_{2,i,j}^{m,n}(t))^{\ast
}a_{i,j+1}^{n\dagger }a_{i,j}^{m}-h_{2,i,j}^{m,n}(t)a_{i,j}^{m\dagger
}a_{i,j+1}^{n}\right] .
\end{eqnarray}
The indices $m$ and $n$ are gauge group indices, and $h_{1,i,j}^{m,n}(t)$
and $h_{2,i,j}^{m,n}(t)$ are background gauge fields. \ We will consider a
simplified version of this system, where 
\begin{eqnarray}
H_{1,i,j}(t) &=&-h_{1,i,j}(t)\left( a_{i+1,j}^{\dagger
}a_{i,j}+a_{i,j}^{\dagger }a_{i+1,j}\right) , \\
H_{2,i,j}(t) &=&-h_{2,i,j}(t)\left( a_{i,j+1}^{\dagger
}a_{i,j}+a_{i,j}^{\dagger }a_{i,j+1}\right) ,
\end{eqnarray}
and $h_{1,i,j}(t)$ and $h_{2,i,j}(t)$ are real valued functions. \ Let us
consider what happens when $h_{1,i,j}(t)$ and $h_{2,i,j}(t)$ are subject to
Gaussian fluctuations. \ Let us define the Gaussian-integrated exponentials
of the Hamiltonian, 
\begin{eqnarray}
A_{1,i,j} &=&\int dh_{1,i,j}(t)\;\exp \left[ -dt\cdot H_{1,i,j}(t)\right]
\exp \left[ -dt\cdot 
{\textstyle{1 \over 8J}}%
(h_{1,i,j}(t)-h)^{2}\right] , \\
A_{2,i,j} &=&\int dh_{2,i,j}(t)\;\exp \left[ -dt\cdot H_{2,i,j}(t)\right]
\exp \left[ -dt\cdot 
{\textstyle{1 \over 8J}}%
(h_{2,i,j}(t)-h)^{2}\right] .
\end{eqnarray}
We find 
\begin{eqnarray}
A_{1,i,j} &\varpropto &\exp \left[ dt\cdot h\left( a_{i+1,j}^{\dagger
}a_{i,j}+a_{i,j}^{\dagger }a_{i+1,j}\right) +dt\cdot 2J\left(
a_{i+1,j}^{\dagger }a_{i,j}+a_{i,j}^{\dagger }a_{i+1,j}\right) ^{2}\right] ,
\\
A_{2,i,j} &\varpropto &\exp \left[ dt\cdot h\left( a_{i,j+1}^{\dagger
}a_{i,j}+a_{i,j}^{\dagger }a_{i,j+1}\right) +dt\cdot 2J\left(
a_{i,j+1}^{\dagger }a_{i,j}+a_{i,j}^{\dagger }a_{i,j+1}\right) ^{2}\right] .
\end{eqnarray}
We note that 
\begin{eqnarray}
\left( a_{i+1,j}^{\dagger }a_{i,j}+a_{i,j}^{\dagger }a_{i+1,j}\right) ^{2}
&=&%
{\textstyle{1 \over 2}}%
\left( -s_{i+1,j}s_{i,j}+1\right) , \\
\left( a_{i,j+1}^{\dagger }a_{i,j}+a_{i,j}^{\dagger }a_{i,j+1}\right) ^{2}
&=&%
{\textstyle{1 \over 2}}%
\left( -s_{i,j}s_{i,j+1}+1\right) ,
\end{eqnarray}
and so the effect of such fluctuations is exactly modelled by an effective
Hamiltonian of the form (\ref{model}) for $J>0$. \ In the staggered
formalism, fermion spin states are staggered at odd and even lattice sites,
and the onset of antiferromagnetic order corresponds with the spontaneous
breaking of chiral symmetry.

The sign problem makes it difficult to study the fermionic Ising model at
large volumes. \ There is no existing method that can handle this model with
explicit fermionic degrees of freedom. \ For example near the transition
temperature for $h/J=2.00$ on a $20\times 20$ lattice we find 
\mbox{$<$}%
Sign%
\mbox{$>$}%
$\sim 10^{-4}$. \ Given the significant computational difficulties we would
like to see if the zone method could be effective for explicit fermion
simulations near the critical temperature. \ One might expect the zone
method to be useful since all of the physics of the $h=0$ Ising model is
already contained in $W_{\emptyset }$, the logarithm of the restricted
partition function with no worldline permutations. \ For $h\neq 0$ the
contributions due to fermion permutations can be included in a controlled
manner by considering successively larger permutation zones.

Let us define the spontaneous staggered magnetization $\left\langle
S\right\rangle $ as 
\begin{equation}
\left\langle S\right\rangle =(-1)^{i+j}\left\langle s_{i,j}\right\rangle
=(-1)^{i+j}\left\langle 2n_{i,j}-1\right\rangle .
\end{equation}
We will compute $\left\langle S\right\rangle $ for various temperatures and
values $h/J$ while adding a small external bias 
\[
H\rightarrow H-m\sum_{i,j}(-1)^{i+j}s_{i,j}.
\]
\ We compute $\left\langle S\right\rangle $ by linearly extrapolating
results for different permutation zone sizes. \ For $h/J\leq 1,$ the sign
problem is significant but still manageable on a $20\times 20$ lattice near
the critical point. \ For $h/J>1$, full simulations on a $20\times 20$
lattice are not practical due to the sign. \ For $h/J=1.50$ we use
permutation zones as large as $10\times 10$, and for $h/J=2.00$ we
extrapolate using zones up to $6\times 6$. \ In Fig. 4 we show the zone
extrapolation for the staggered magnetization $\left\langle S\right\rangle $
at $h/J=1.50$, $T/J=2.00$, and $m/J=0.0256$ on a $20\times 20$ lattice. \ In
Fig. 5 we plot the zone-extrapolated results as a function of $T/J$ near the
critical temperature for $h/J=1.50$, $m/J=0.0256$. \ Since the hopping
interaction increases the disorder of the system, we expect the critical
temperature to decrease as $h$ increases. \ Noting that the $h=0$ Ising
model has a critical temperature

\begin{equation}
T_{c}^{\text{Ising}}/J=%
{\textstyle{2 \over \ln (1+\sqrt{2})}}%
\approx 2.269,
\end{equation}
we see that the data in Fig. 5 does in fact show a decrease in the critical
temperature. \ We also find a critical exponent of $\beta =0.10(2)$, which
is consistent with that of the Ising model, $\beta _{\text{Ising}}=0.125.$

In Fig. 6 we show the deviation of the critical temperature from $T_{c}^{%
\text{Ising}}$. \ In our plot we have included errors due to finite size
effects, which we found by recalculating results using $10\times 10$ and $%
16\times 16$ lattice sizes and several different values for $m/J$. \ We note
that the deviation of the critical temperature appears to be quadratic in $%
h/J,$ with possibly a small negative cubic contribution$.$ \ It would be
interesting to carry these simulations further to see how the curve behaves
for larger values of $h/J$. \ About 2000 CPU hours on an IBM SP3 were used
to collect the data for this study.

\section{Summary}

We have presented a promising new approach to fermionic simulations at
finite temperatures. \ We have demonstrated that the exchange of identical
fermions is short-ranged and has a maximum range of $\sqrt{2(1-f)\beta h}$
lattice sites$,$ where $\beta $ is the inverse temperature, $h$ is the
hopping parameter, and $f$ is the filling fraction. \ We have introduced the
notion of permutation zones, special regions of the lattice where identical
fermions may interchange and outside of which they may not. $\ $Using
successively larger zones, one can extrapolate to obtain thermodynamic
observables. \ We have demonstrated the method using two different examples:
\ the average energy for a 2D free fermion and the spontaneous staggered
spin magnetization for the 2D fermionic Ising model.

The methods presented here are relatively easy to implement. \ Correlation
functions can be calculated using worm algorithms \cite{prokofev}, which are
generalizations of the loop algorithm. \ Although the full applicability of
the zone method still needs to be determined, the method and its future
generalizations may have some impact on several problems in strongly
correlated solid-state systems and lattice gauge theory. \ In this paper we
have focused on simulations with explicit fermion degrees of freedom. \ The
extension of the zone method to pseudofermion algorithms is currently being
studied and will presented in a future publication.

\acknowledgements
The author thanks Shailesh Chandrasekharan, Hans Evertz, Pieter Maris,
Nikolai Prokof'ev, and Nathan Salwen for discussions and the North Carolina
Supercomputing Center for supercomputing time and support.

\begin{figure}[htbp]
\vspace{2mm}
\begin{center}
\epsfig{figure=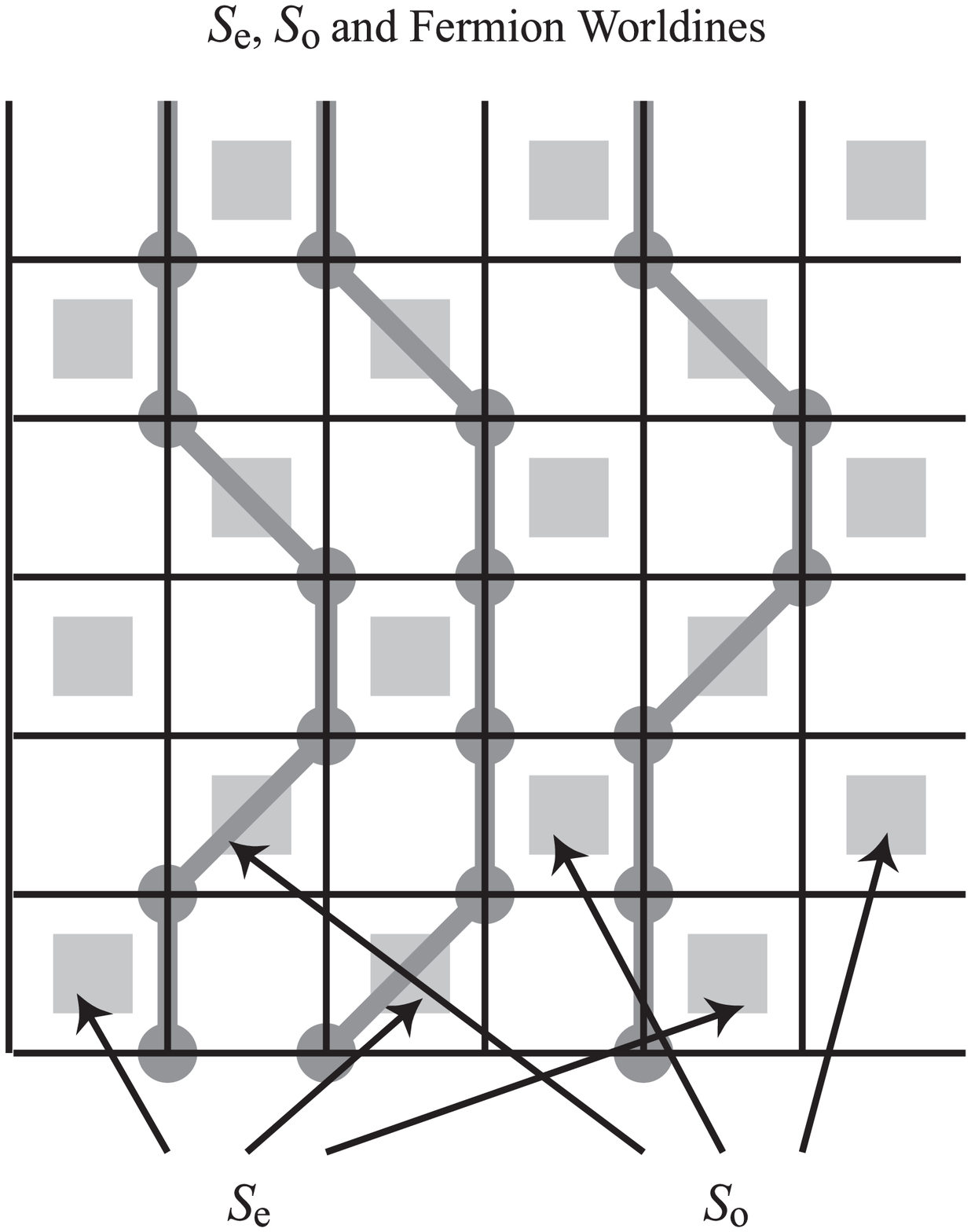,width=70mm}
\end{center}
\vspace{2mm}
\caption{Typical worldline configuration for the one-dimensional system.}
\end{figure}

\begin{figure}[htbp]
\vspace{2mm}
\begin{center}
\epsfig{figure=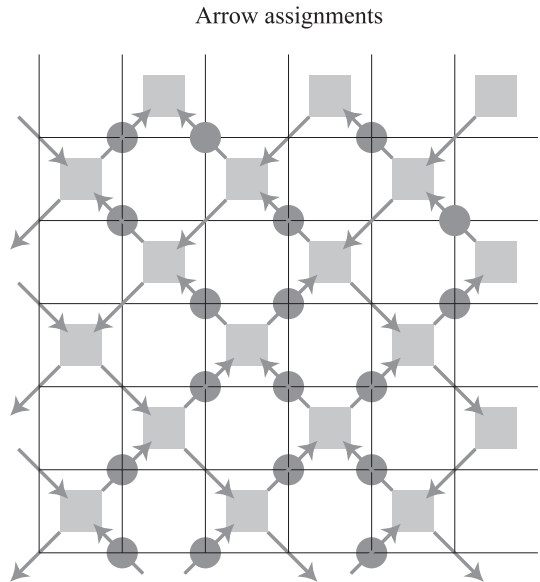,width=70mm}
\end{center}
\vspace{2mm}
\caption{Upward/downward arrows are drawn at each occupied/unoccupied site.  
For each plaquette the number of inward-pointing arrows 
equals the number of outward-pointing arrows.}
\end{figure}

\begin{figure}[htbp]
\vspace{2mm}
\begin{center}
\epsfig{figure=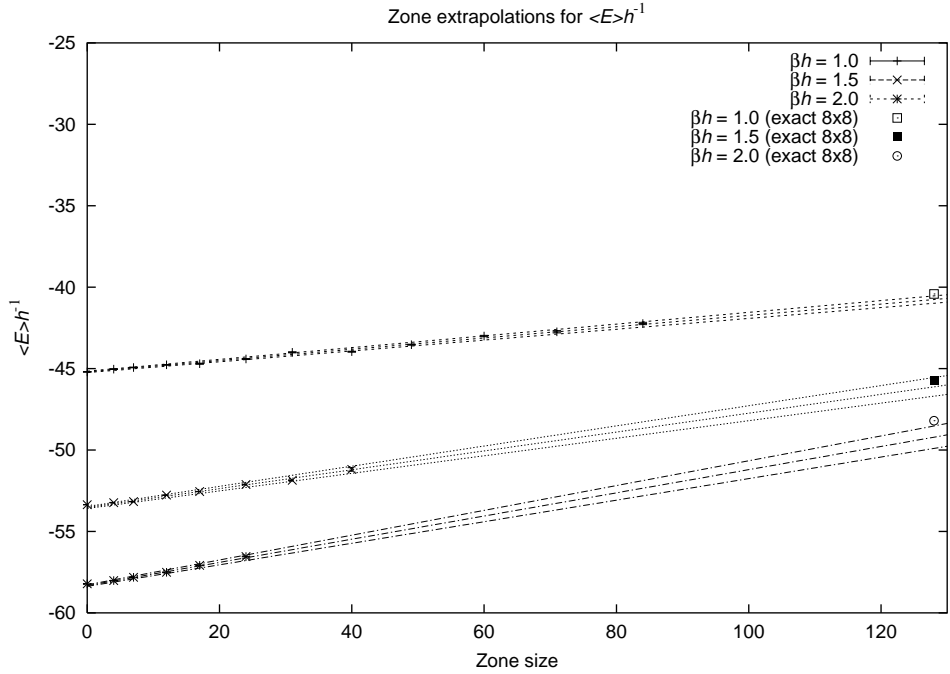,angle=-90,width=130mm}
\end{center}
\vspace{2mm}
\caption{Average energy at $\protect\beta h=1.0,1.5,2.0$ for free
fermions on an $8\times 8$ lattice$.$}
\end{figure}

\begin{figure}[htbp]
\vspace{2mm}
\begin{center}
\epsfig{figure=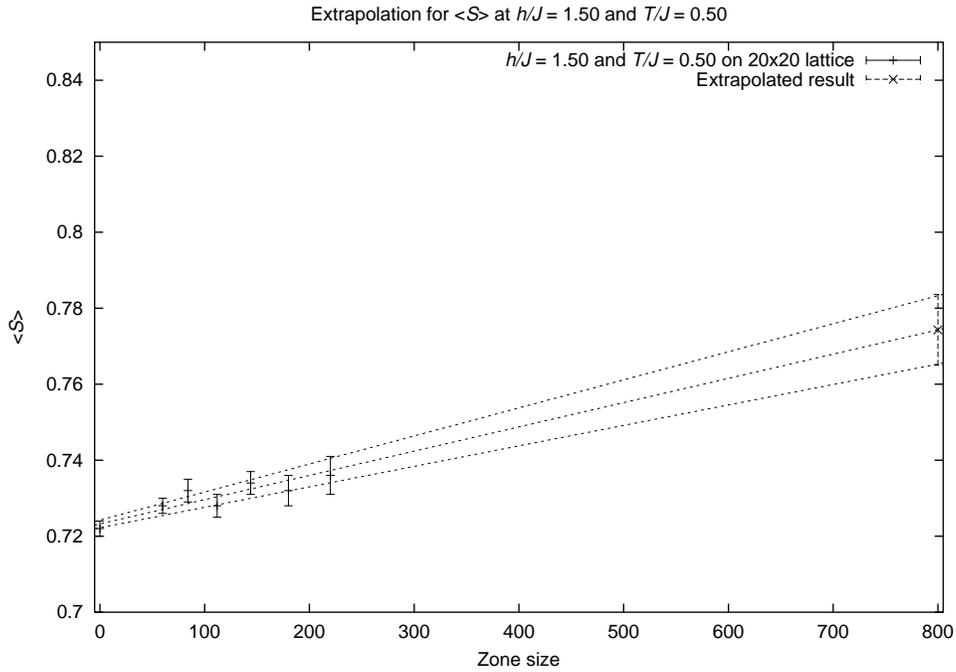,angle=-90,width=130mm}
\end{center}
\vspace{2mm}
\caption{Zone extrapolation for the staggered magnetization 
$\left\langle S\right\rangle$ 
for the 2D fermionic Ising model at $h/J = 1.50, T/J = 2.00, m/J = 0.0256.$}
\end{figure}

\begin{figure}[htbp]
\vspace{2mm}
\begin{center}
\epsfig{figure=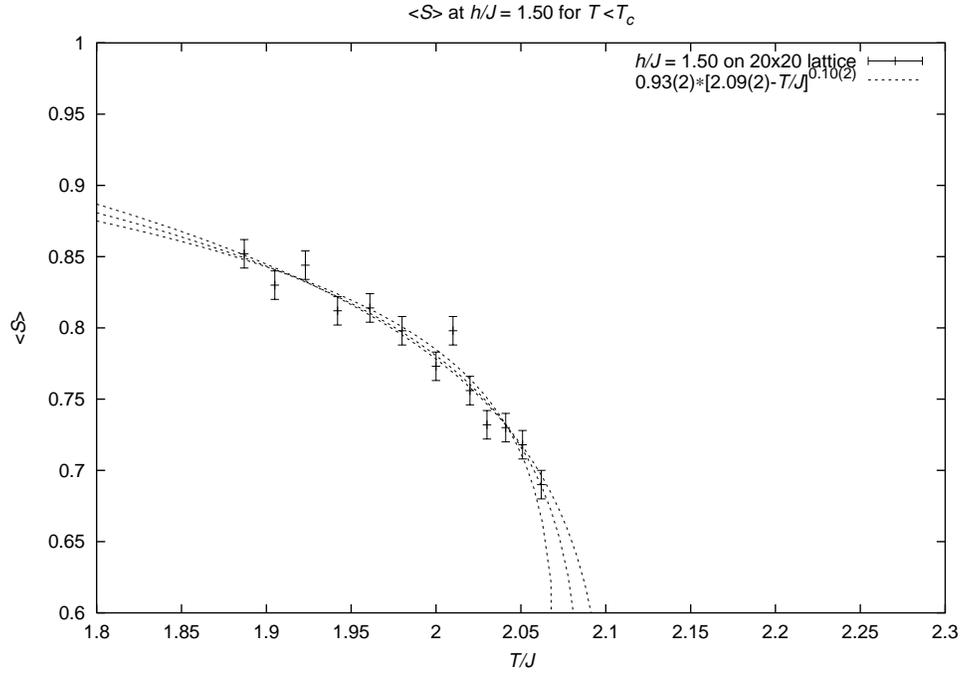,angle=-90,width=130mm}
\end{center}
\vspace{2mm}
\caption{Staggered magnetization $\left\langle S\right\rangle$ for the 
2D fermionic Ising model near the critical point for $h/J = 1.50, m/J = 0.0256$.}
\end{figure}

\begin{figure}[htbp]
\vspace{2mm}
\begin{center}
\epsfig{figure=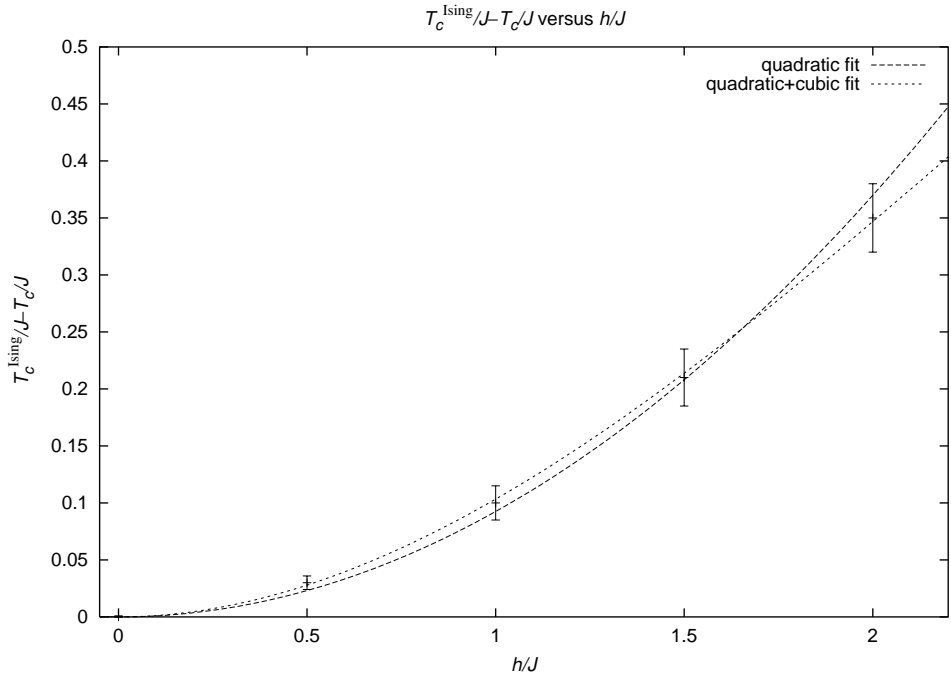,angle=-90,width=130mm}
\end{center}
\vspace{2mm}
\caption{Change in the critical temperature of the 2D fermionic Ising 
model as a function of $h/J$.}
\end{figure}

\end{document}